%% file: InclusiveCrossSection2011.tex
\documentclass[final,5p,times,twocolumn]{elsarticle}

\usepackage{amssymb}

\usepackage{hyperref}
\usepackage{graphicx}
\DeclareGraphicsExtensions{.pdf}
\usepackage[percent]{overpic}
\usepackage{atlasphysics}
\usepackage{amsmath}

\input{definitions}

\journal{Physics Letters B}

\begin{document}

\begin{frontmatter}

\title{\vspace{-0.5cm} \flushright{\normalsize{CERN-PH-EP-2011-115}} \\ \center{
  Measurement of the inclusive isolated prompt photon cross-section \\
  in $pp$ collisions at $\sqrt{s}= 7\TeV$ 
  using $35~\ipb$ of ATLAS data
}}
\author{ATLAS Collaboration}

\begin{abstract}
  A measurement of the differential cross-section for the inclusive
  production of isolated prompt photons in $pp$ collisions at a
  center-of-mass energy $\sqrt{s} = 7\TeV$ is presented.  The
  measurement covers the pseudorapidity ranges $|\eta|<1.37$ and
  $1.52\leq|\eta|<2.37$ in the transverse energy range $45\leq\ET<400
  \GeV$. The results are based on an integrated luminosity of 35~\ipb,
  collected with the ATLAS detector at the LHC.  
  The yields of the signal photons are measured using a data-driven
  technique, based on the observed distribution of the hadronic
  energy in a narrow cone around the photon candidate and the
  photon selection criteria.
  The results are compared with next-to-leading order perturbative QCD
  calculations and found to be in good agreement over four orders of
  magnitude in cross-section.
\end{abstract}


\end{frontmatter}

The production of prompt photons at hadron colliders provides means
for testing perturbative QCD predictions~\cite{QCD_promptphoton_NLO},
providing a colorless probe of the hard scattering process. The
measurement of the inclusive production of prompt photons could be
used to constrain the parton distribution functions; in particular it
is sensitive to the gluon content of the
proton~\cite{promptphoton_and_gluon_pdf} through the
$qg~\rightarrow~q\gamma$ sub-process, which at leading-order dominates
the inclusive prompt photon cross-section at the LHC.

ATLAS has recently published a measurement of the inclusive photon
cross-section in $pp$ collisions at $\sqrt{s}= 7\TeV$ using an
integrated luminosity of 880~\inb~\cite{InclusivePaper2010}; a similar
measurement has been performed by the CMS
collaboration~\cite{PhysRevLett.106.082001} using an integrated
luminosity of 2.9~\ipb. 
Analogous measurements have been perfomed in $p\bar{p}$ collisions at
a lower center of mass at the Tevatron
\cite{PhysRevD.80.111106,Abazov2006151}, and in deep inelastic $ep$
scattering at HERA \cite{epjc-s10052-008-0541-6,Chekanov:2009dq}.
This letter presents the measurement of the differential
production cross-section of isolated prompt photons with transverse
energies $\et$ above $45\GeV$ using 34.6~$\pm$~1.2~$\ipb$ of
$pp$ collision data at $\sqrt{s}= 7\TeV$ collected in 2010.
Isolated prompt photons in the pseudorapidity ranges $|\eta|<0.6$, $0.6\leq
|\eta|<1.37$, $1.52\leq |\eta| < 1.81$ and $1.81\leq
|\eta| < 2.37$ are studied~\cite{AtlasCoords}.

In the following, all photons produced in $pp$ collisions and not
coming from hadron decays are considered as \emph{prompt}: they
include both \emph{direct} photons, which originate from the hard
sub-process, and \emph{fragmentation} photons, which are the result of
the fragmentation of a colored high-\pt\
parton~\cite{jetphox1,jetphox2}. \emph{Isolated} photons are
considered: from a theoretical perspective, photons are isolated if
the transverse energy $\ET^{\rm iso}$, within a cone of radius 
$R~=~\sqrt{(\Delta\eta)^2+(\Delta\phi)^2}$~=~0.4 centered around the
photon direction in the pseudorapidity $(\eta)$ and azimuthal angle
$(\phi)$ plane~\cite{AtlasCoords}, is smaller than
$\ET^{\rm cut}$. In \jetphox{}~\cite{jetphox1}, used for next-to-leading order 
(NLO) calculations, $\ET^{\rm iso}$ is calculated from all partons.
Similarly, a corresponding isolation prescription is
applied experimentally on the reconstructed objects, based on the
energy reconstructed in an $R$~=~0.4 cone around the photon candidate,
corrected for the effects associated with: the energy of the photon
candidate itself, the underlying event and the collision
pile-up~\cite{InclusivePaper2010}. The main background to these
isolated prompt photons is composed of photons from decays of light
neutral mesons, such as the $\pi^{0}$ or $\eta$.

Photons are detected in ATLAS by a lead-liquid Argon sampling
electromagnetic calorimeter (ECAL) with an accordion geometry, divided
into a barrel section covering the pseudorapidity region $|\eta|<
1.475$ and two end-cap sections covering the pseudorapidity regions
$1.375<|\eta|<3.2$. It consists of three longitudinal layers. The
first layer has a high granularity along the $\eta$ direction (between
0.003 and 0.006 depending on $\eta$, with the exception of the regions
$1.4<|\eta|<1.5$ and $|\eta|>2.4$), sufficient to provide an
event-by-event discrimination between single photon showers and
showers coming from a $\pi^0$ decay. The second layer has a
granularity of $0.025\times0.025$ in $\eta\times\phi$. A third layer
is used to correct for the leakage beyond the electromagnetic
calorimeter for high-energy showers, while in front of the accordion calorimeter a thin
presampler layer, covering the pseudorapidity interval $|\eta|<1.8$,
is used to correct for the energy absorbed before the calorimeter.

The ECAL energy resolution is parametrized as $\sigma(E)/E =
a/\sqrt{E~{\rm (GeV)}} \oplus c$ with the largest contribution coming
from the sampling term $a$, corresponding to approximately $10\%$~($20\%$) in the
barrel~(endcap) region. For energies above $200\GeV$ the global constant term $c$,
estimated to be $(1.2\pm{}0.6)\%$~(($1.8\pm{}0.6)\%$) in the
barrel~(endcap) for the 2010 data, starts to
dominate~\cite{ATLAS-PERF-2010-04-001}.
In front of the electromagnetic calorimeter the inner detector allows the reconstruction of
tracks from the primary $pp$ collision point and also from
secondary vertices, permitting an efficient reconstruction of photon
conversions in the beam pipe and inner detector up to a radius of
$\sim$~80~cm.
Further details of the inner detector, the electromagnetic calorimeter and the whole ATLAS detector 
are documented in Ref.~\cite{ATLAS_detector}.

Event samples simulated with \pythia{}~6.4.21 \cite{pythia} are used to study the characteristics of signal and background events.
To estimate systematic uncertainties related to the choice of the event generator and the parton shower model, alternative samples are generated with
\herwig{}~6.5 \cite{Herwig}.
Events used in this analysis are triggered using a single-photon
trigger with a nominal transverse energy threshold of
40~\GeV.
The trigger efficiency, \efftrig, is measured using a
bootstrap method to be $(99.4^{+0.6}_{-0.2})\%$ for prompt photon
candidates with $\et > 45 \GeV$ passing the selection criteria
presented below. The same trigger condition was used for the whole
dataset, even though the mean number of events per collision rose from
$<1$ to $\sim3$ as the instantaneous luminosity increased during 2010.
Collision candidates are selected by requiring a primary vertex with
at least three associated charged particle tracks, consistent with the
beam interaction region. The total number of selected events in data
after these requirements is almost 1.7 million, with a negligible amount 
of non-collision background.

Photon candidates are formed from clusters of energy deposits reconstructed in
the electromagnetic calorimeter~\cite{ATLAS_CSC}.  Clusters without matching tracks are
classified as \emph{unconverted} photon candidates. The presence of one
or two tracks coming from a conversion vertex is used to distinguish
\emph{converted} photons from electrons. Converted photon clusters are rebuilt with a wider size in
$\phi$, to account for the opening angle between the conversion products
due to the magnetic field. A specific energy
calibration~\cite{ATLAS_CSC} is then applied separately for converted
and unconverted photon candidates to account for energy loss in front of the ECAL
 and both lateral and longitudinal leakage. Photon clusters are removed if their barycenter lies
in the transition between the barrel and endcap regions of the electromagnetic calorimeter, corresponding to $1.37<|\eta|<1.52$, 
where larger uncertainties related to the efficiency measurement are expected.
Clusters containing cells overlapping with the small number
of regions with problematic calorimeter readout or with very noisy
cells are also removed.  Over 0.8 million photon candidates with
$\et>45\GeV$ remain in the data sample.

A measurement of the transverse isolation energy \Etiso\ is associated
with each photon candidate, computed by summing the calorimeter energy
in a cone of $R~=~0.4$ around the candidate, as detailed in
Ref.~\cite{InclusivePaper2010}.  Corrections to this isolation energy
are derived from simulation to remove the energy of the photon itself
that leaks into the isolation cone. An event-by-event
correction~\cite{Cacciari:area,Cacciari:UE} is applied to subtract the
estimated contributions from the underlying event and in-time
pileup (i.e.\ from additional proton-proton interactions). The correction to \Etiso\ is 
typically 900~\MeV. After this
subtraction, the remaining fluctuations are dominated by electronic
noise from the calorimeter measurement. The effect of the out-of-time
pileup, associated with collisions taking place in previous
bunch-crossings, is found to be minimal (i.e.\ shifts of 200
MeV at most, towards lower isolation energies).  The corrections
mentioned above allow \Etiso\ to be directly compared to
parton-level theoretical predictions.

All photon candidates having reconstructed isolation energy $<3$~\GeV\
are considered as experimentally isolated. This definition
is similar to applying a 4~\GeV\ cut on the particle-level
isolation, defined as the transverse energy of all stable particles in
a cone of radius $R=0.4$ around the photon direction (with the underlying event removed as before). The small
difference between the two, caused by noise and other detector effects, is taken into account in the uncertainties
associated with the photon reconstruction efficiency \effreco\ discussed
below. The particle-level isolation can in turn be related to the
parton-level isolation in \jetphox{} that is used for
the NLO predictions. 
The efficiency of the isolation criteria is found to be
similar (i.e.\ within a few percent) at both the particle-level and the parton-level for 
simulated photons passing the selection described below.

As in Ref.~\cite{InclusivePaper2010}, the reconstruction and
preselection efficiency \effreco\ is computed from simulated prompt
photons as a function of the true photon $\et$.  It is defined as the
ratio between the number of photons reconstructed in a given $|\eta|$
interval with reconstructed $\Etiso<3\GeV$, and the total number of
true prompt photons with true pseudorapidity in the same $|\eta|$
interval, and with particle-level transverse isolation energy $<4$~\GeV.  
The estimated \effreco\ for photons with 45 $< \et < 400\GeV$ is 
$\sim85\%$~(75\%) 
in the barrel~(endcap) region.
  The main inefficiency ($\sim10\%$) is due to
the acceptance loss originating from a few inoperative optical links in the
calorimeter readout. A similar reduction is caused by the isolation
requirement in the pseudorapidity region $1.52\leq|\eta|<1.81$ where
the calorimetric isolation suffers from larger detector effects.  The
systematic uncertainty on \effreco\ associated with the experimental
isolation requirement is evaluated from the prompt photon simulation
by varying the value of the isolation criterion by the average
difference ($\sim500\MeV$) observed for electrons from $W\to e\nu$
events in data and simulation. The estimated uncertainty varies between 3$\%$ 
and 4$\%$ depending on $\eta$. The uncertainty associated with the imperfect
knowledge of the material in front of the ECAL is estimated by
comparing the expected efficiencies in a sample simulated with the
nominal ATLAS setup, and one with increased material. It varies
between 1\% and 2.5\%, depending on \eta.

Shape variables computed from the lateral and longitudinal energy
profiles of the shower in the calorimeters are used to discriminate
signal from background~\cite{ATLAS_CSC,ATL-PHYS-PUB-2011-007}. As
detailed in Ref.~\cite{InclusivePaper2010}, selection criteria on
these variables, optimized independently for unconverted and converted
photons, are applied to reconstructed photon candidates. The requirements
on these variables are applied in stages resulting in \emph{tight} candidates:
firstly jets are removed whilst still keeping a high photon efficiency and then
secondly wide or closely spaced showers (i.e.\ those consistent with jets or
meson decays) are rejected. The selection
criteria have been revised to minimize the systematics on the efficiency
extraction, especially in the region \etafour.  The photon
identification efficiency \effid\ is computed from simulation as a
function of transverse energy in each pseudorapidity region. It is
defined as the efficiency for reconstructed (true) prompt photons,
with measured $\Etiso<3\GeV$, to pass the identification criteria
mentioned above.

Following the same method as Ref.~\cite{InclusivePaper2010}, the
value of \effid\ is determined after correcting the simulated shower
shapes for the observed average differences with respect to data.  In
the present analysis, however, the corrections are estimated for
unconverted and converted photons separately. This helps to reduce the
systematic uncertainties associated with the correction procedure.
The value of \effid\ varies from 90\% to 97\%, depending on $\eta$ and increasing
with $\et$. The systematic uncertainty on \effid\ is also $\eta$
dependent, ranging from 1.5\% to 3\%, with contributions from: detector
simulation; background contamination; (un)converted photon
misclassification; direct/fragmentation photon fraction; the choice of
different Monte Carlo generators (MC). These uncertainties affect the
reconstruction and identification efficiencies in a correlated way,
and are treated as such in their combination. After applying the 
isolation criterion and the tight selection on the shape variables, 
almost 173,000 photon candidates remain in the data sample.

\begin{figure}
  \centering
  \includegraphics[width=\columnwidth]{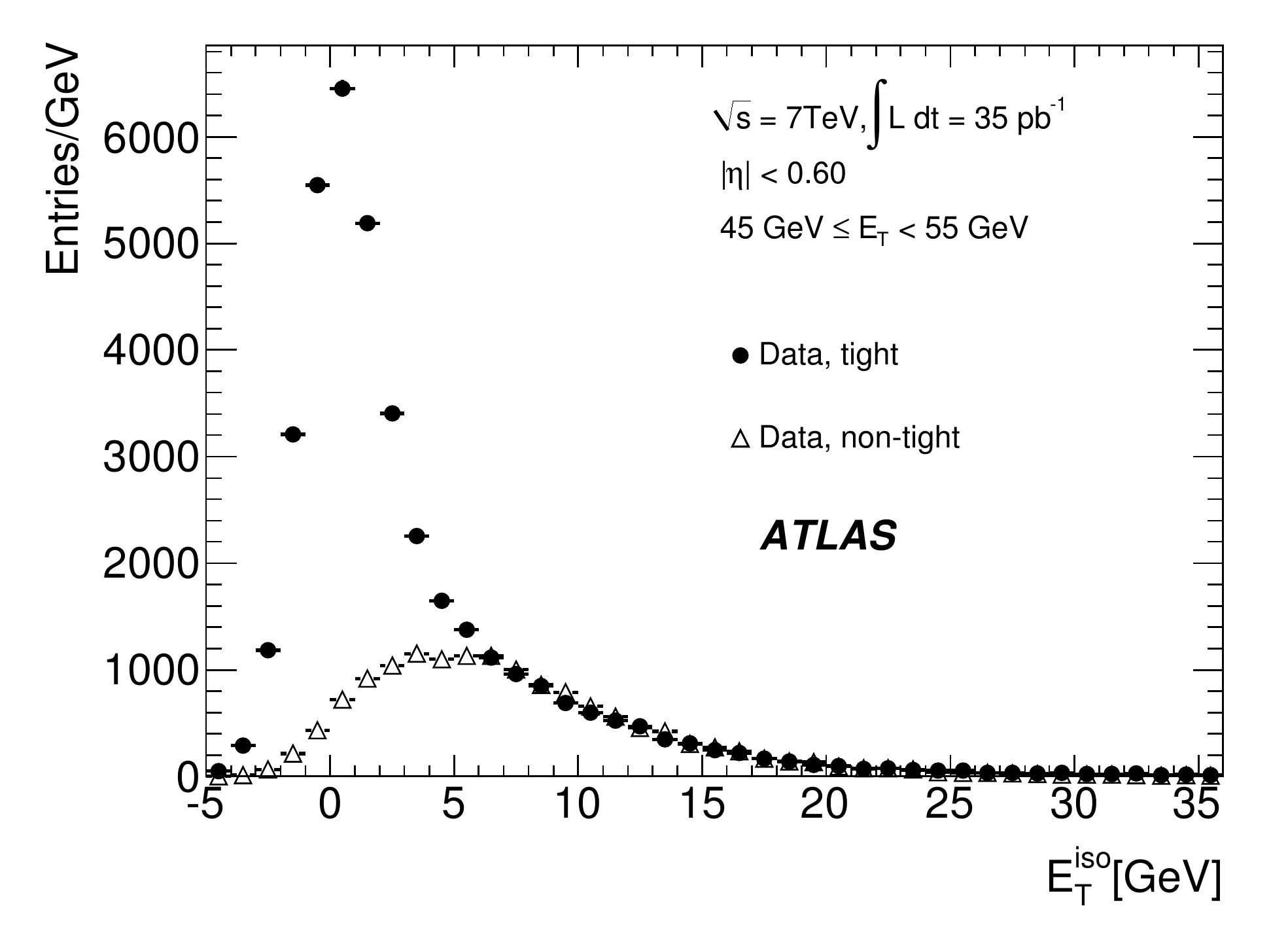}
  \caption{Distributions of \Etiso\ for photon candidates with $45\GeV<\et<55\GeV$ in
    $|\eta|<0.6$ passing the tight (solid dots) and non-tight (open triangles) shower-shape-based selection criteria.
    The non-tight distribution is normalized to the tight distribution for $\Etiso>5$~\GeV\
    (non-isolated region), where the signal contamination is fairly small.}
  \label{fig:2DS_ptbin1_etabin1}
\end{figure}

As in Ref.~\cite{InclusivePaper2010}, a two-dimensional-sideband
method is used to estimate the background contribution from data and
to measure the prompt photon signal yield. The two dimensions are the
transverse isolation energy \Etiso\ and the quality of the photon,
defined by whether or not it passes the shower shape identification
criteria.  On the isolation axis, the signal region contains photon
candidates with $\Etiso\ <3$ GeV, while the sideband region contains
\emph{non-isolated} photon candidates with $\Etiso\ >5$ GeV.  On the other
axis, the signal photon candidates are required to pass the tight
identification criteria (\emph{tight} candidates). Those failing the
tight criteria but passing a background-enriching subset of these criteria
(\emph{non-tight} candidates) are contained in the sideband.  A typical
distribution of \Etiso\ for both tight and non-tight data is shown in
Fig.~\ref{fig:2DS_ptbin1_etabin1} for photon candidates with
$45\GeV<\et<55\GeV$ in $|\eta|<0.6$. The non-tight distribution is
normalized to the tight one above 5~\GeV\, where a only small signal
contamination is expected.

Corrections for the signal contamination in the background control
regions are computed using prompt photon Monte Carlo samples. For the
tight isolated signal leaking into the non-isolated region, these are
as large as 17\% at high $\et$. Smaller leakages of up to 6\% are
expected for the other two background control regions. The purity of
isolated prompt photons measured with this method increases with $\et$
from 91\% at $\et=45\GeV$ to close to 100\% at $\et>200\GeV$.

The main contributions to the uncertainty on the yields come from the 
fragmentation fraction ($\lesssim$ 8\%),
estimated by conservatively varying the fraction from 0 to 100$\%$ in the signal sample, 
and pileup ($5\%$, with fluctuations up to $8\%$ for \etathree),
estimated by increasing the correction to \Etiso\ by $50\%$ both in data and simulation.
This scaling of the correction minimizes the residual dependency of the isolation 
on the number of primary vertices (i.e.\ pile-up) in data. 
The other contributions to the uncertainty are: correlated background in the two-dimensional-sideband regions ($\lesssim$5\% barrel and $\lesssim$10\% endcap, $\et$ dependent),
definition of the two-dimensional-sideband regions ($\lesssim$ 5\% non-tight and $1\%$ non-isolated),
photon energy scale ($2$-$8\%$, $\eta$ dependent),
slightly narrower showers in simulation than in data ($2\text{-}5\%$, $\eta$ and $\et$ dependent),
isolation shower leakage corrections ($1\text{-}5\%$),
Monte Carlo generator ($2\%$),
material effects ($<1\%$),
and prompt electron misidentification ($\sim0.5\%$, varying with $\et$). 
Globally, the uncertainties on the photon signal
yields are less than 10$\%$, and decrease with $\et$.


The average differential cross-section
$\left\langle d\sigma_j^{k} / d\ET^{\rm true} \right\rangle$
for the production of isolated prompt photons in a bin $j$ of
$\ET^{\rm true}$ (integrated over one true $|\eta|$ bin $k$) is related to
the signal yield $N^{\gamma,{\rm reco},k}_{i}$ (in the $k$'th $|\eta|$
bin and $i$'th $\et$ bin) by the relationship:

\begin{align}
N^{\gamma,{\rm reco},k}_{i} = &
  \left(   \int{\mathcal{L}dt} \right) 
  \efftrig \varepsilon^{{\rm ID},k}_i \times \nonumber \\
  & 
  \sum_{j}   R_{ij}^k\varepsilon^{{\rm reco},k}_j\ \Delta E_{{\rm T},j}^{\rm
  true}\ \left\langle\frac{d \sigma_j^{k}}{d\ET^{\rm
  true}}\right\rangle 
\end{align}

where $\varepsilon^{{\rm ID},k}_i$ is the average 
identification efficiency and $R_{ij}^k$ is the {\em $\et$ response matrix}.
The elements of $R_{ij}^k$ are
evaluated from the ratio of the true to reconstructed
$\et$ distributions of photon candidates, using simulated samples of isolated prompt photons. 
The migration from one $\et$ bin to another is less than 10\% in most $\et$ and $\eta$ regions. A larger migration of
up to 18\%\ is observed in the region \etathree, where more material is present in front of
the electromagnetic calorimeter. Migrations between $\eta$ bins are neglected given the large bin size and the excellent ECAL $\eta$ resolution.
A singular value decomposition (SVD) ~\cite{Hocker:1996kb} is used to unfold the $\et$ distribution for detector effects.
The regularization of the resulting unfolded
distribution is tuned using simulated events and chosen to be very
loose to avoid a potential bias toward the truth reference
spectrum. The simulation model dependence is tested with pseudo-experiments, using \pythia{} and \herwig{} simulated samples. 
The difference of the unfolded cross-section obtained in both cases is found to be $<3\%$.  The uncertainty associated 
with the ECAL energy resolution is $\sim$ 1\%. The lower and upper $\et$ constraints have
negligible effect on the unfolded spectrum.

\begin{figure*}[p]
  \centering
  \begin{tabular}{cc}
  \begin{overpic}[width=0.5\textwidth]{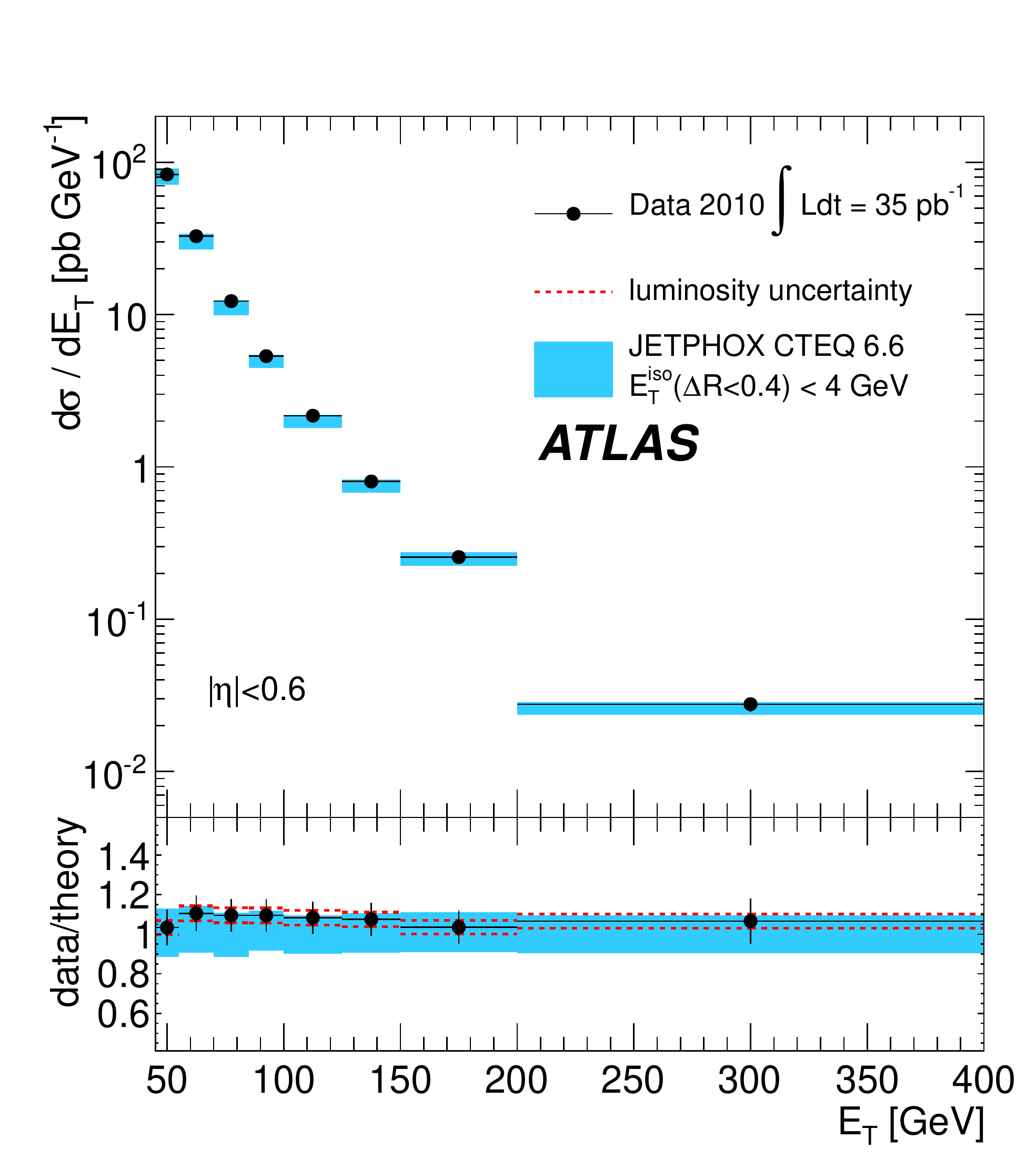}
    \put(17,45){(a)}
  \end{overpic}
&
  \begin{overpic}[width=0.5\textwidth]{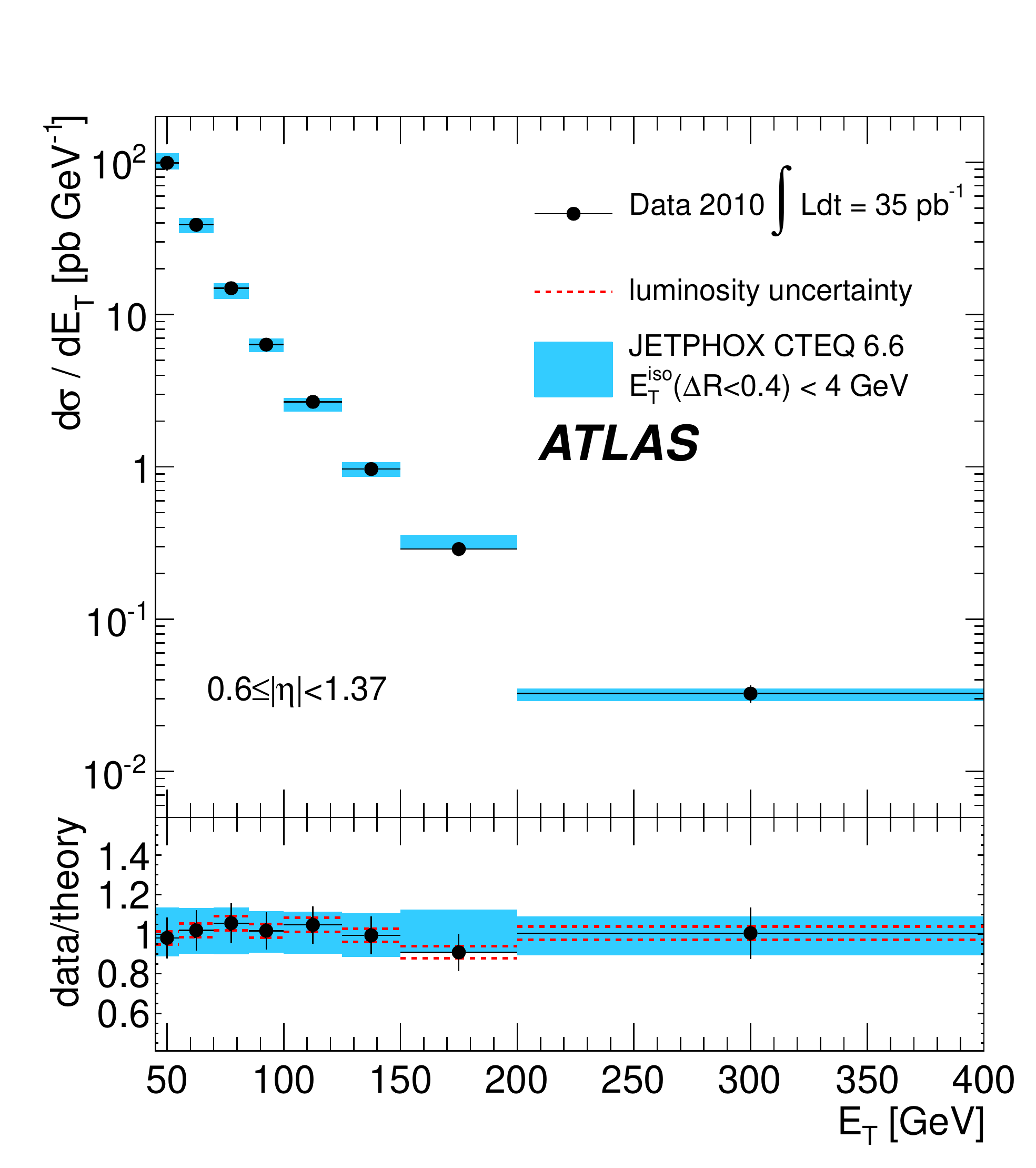}
    \put(17,45){(b)}
  \end{overpic}
\\
  \begin{overpic}[width=0.5\textwidth]{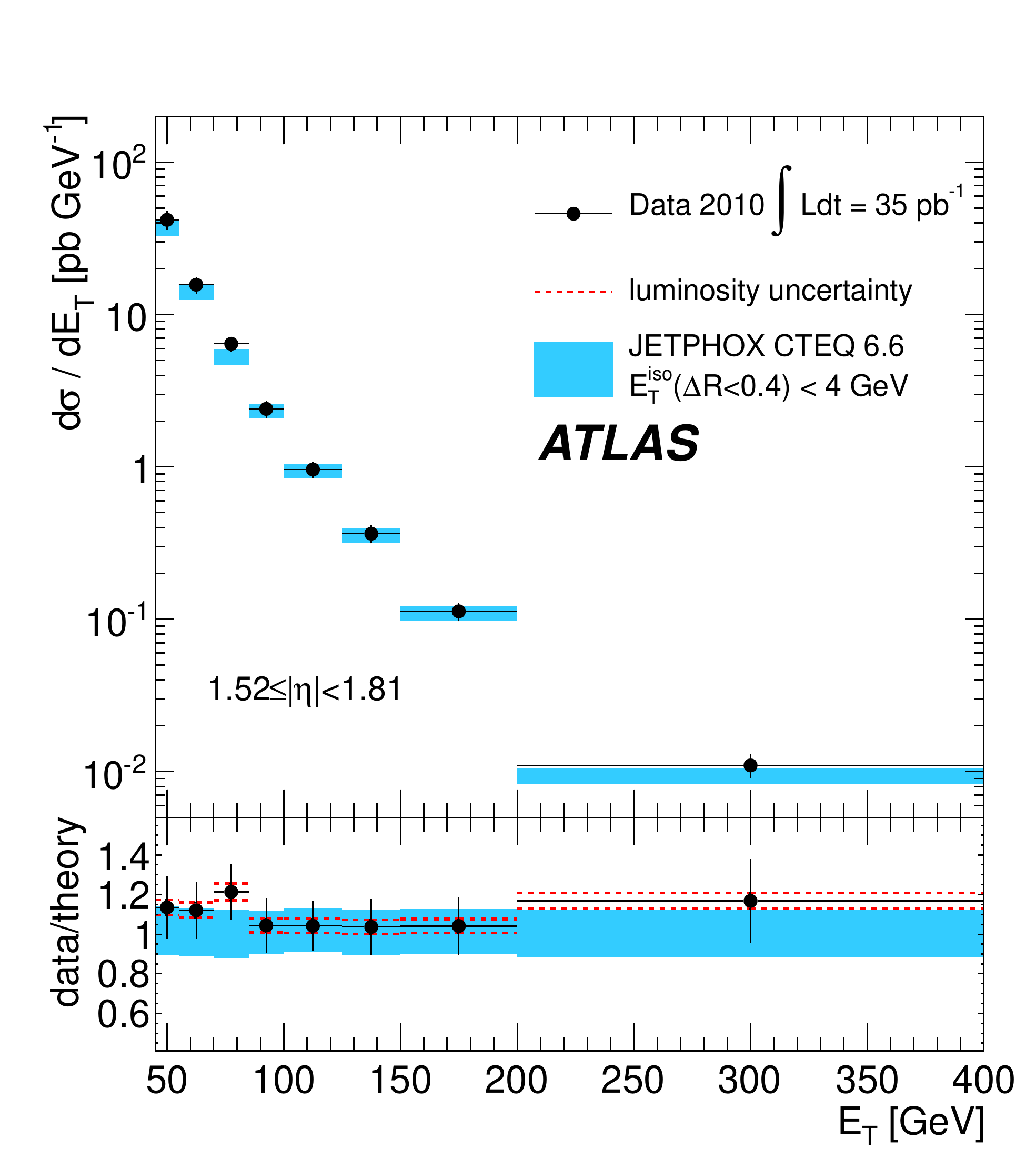}
    \put(17,45){(c)}
  \end{overpic}
&
  \begin{overpic}[width=0.5\textwidth]{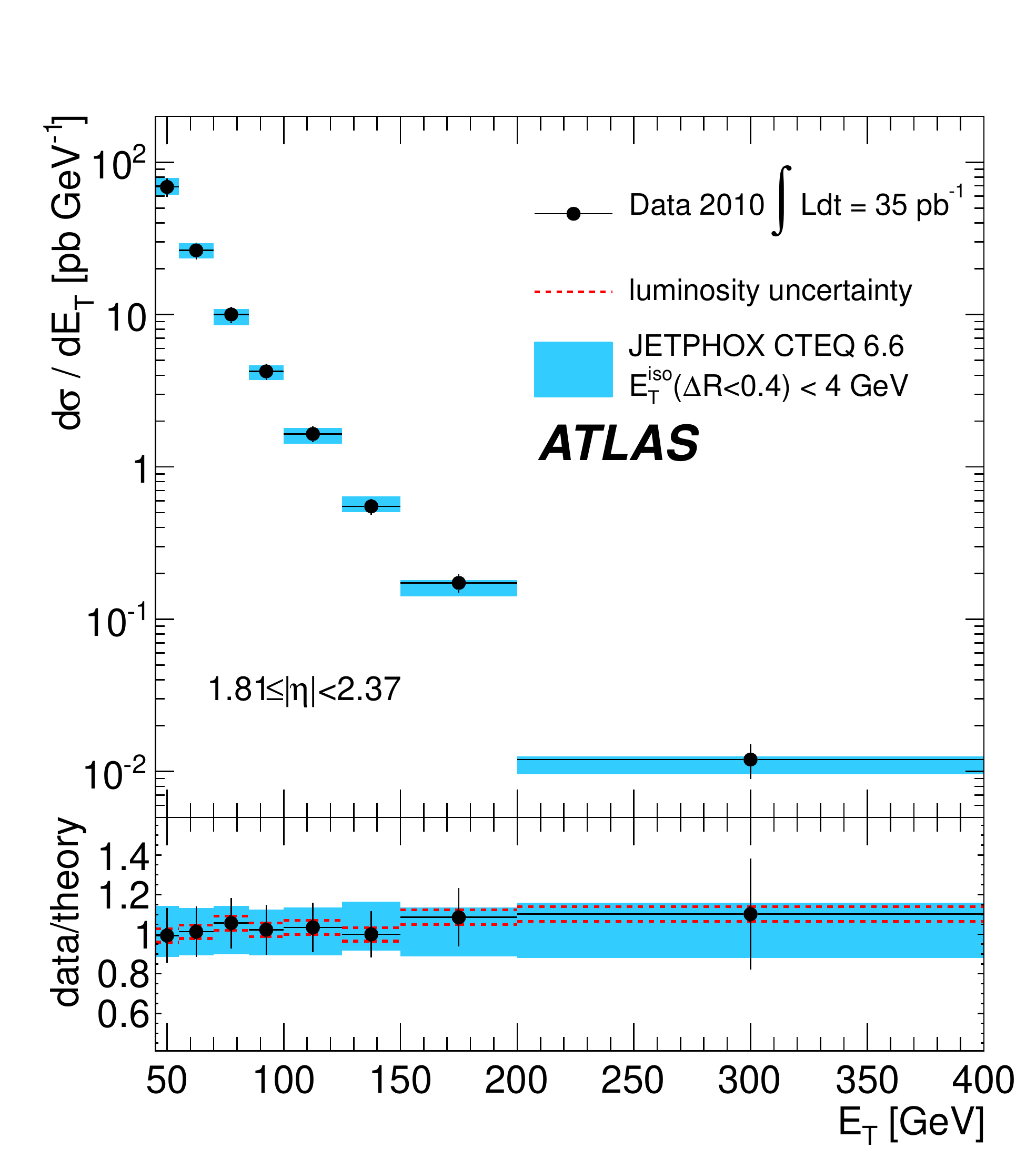}
    \put(17,45){(d)}
  \end{overpic}
  \end{tabular}
  \caption{Measured (dots) and expected (shaded area) inclusive prompt
    photon production cross-sections, and their ratio, as a function
    of the photon $\et$ and in the range (a) $|\eta|<0.6$, (b) $\etatwo$, (c) $\etathree$ and (d) $\etafour$. The data error bars combine the statistical and systematic uncertainties, with the luminosity uncertainty shown separately (dotted bands).}
  \label{fig:xsec_measured_vs_jetphox_2dsideband_etabin000-237}
\end{figure*}

The measured inclusive isolated prompt photon production cross-sections are shown in
Fig. \ref{fig:xsec_measured_vs_jetphox_2dsideband_etabin000-237}. They are
presented as a function of the photon transverse energy, for each of
the four considered pseudorapidity intervals. They are also presented
in tabular form in~\ref{app:CStables}. 
The error bars on the data points represent the combination of the
statistical and systematic uncertainties: systematic uncertainties
dominate over the entire kinematic range considered. The contribution
from the luminosity uncertainty (3.4\%) is shown separately as it
represents a possible global change by a common multiplicative factor.
The data agree with NLO pQCD calculations, obtained with \jetphox{}~1.2.2~\cite{jetphox1}
using the CTEQ 6.6 PDFs~\cite{cteq} and the BFG set II~\cite{Bourhis:2000gs} fragmentation
functions (FF). These predictions are negligibly
affected when using BFG set I instead. The nominal renormalization,
factorization and fragmentation scales are set to the $\et$ of the
photon. Theoretical calculations using MSTW 2008~\cite{mstw} and NNPDF2.0~\cite{nnpdf} PDFs show
a similarly good agreement to data. The central values
obtained with the MSTW 2008 (NNPDF2.0) PDFs are 3 to 5\% (1 to 4\%)
higher than those predicted using the CTEQ 6.6 PDFs.
The total systematic uncertainties on the theoretical
predictions are represented with a solid band. The scale uncertainty 
($\sim10~\%$) is the leading theoretical systematic uncertainty. It is estimated from the
envelope of independent and coherent variations of the three scales, by a
factor of two around the central value, with the renormalization scale
(coherent variation) dominating this envelope at low (high) $\et$,
while the fragmentation scale produces the smallest variation.
The scale error is summed in quadrature with the contributions from
the PDF uncertainty (5\% at 68\% C.L.) and the uncertainty associated with
the choice of the parton-level isolation criterion (2\%).  The same
quantities are also shown in the bottom panels after having been
normalized to the expected NLO pQCD cross-sections.

In conclusion, the inclusive isolated prompt photon production cross-section 
in $pp$ collisions at a center-of-mass energy $\sqrt{s}=7$ TeV
has been measured using 35~\ipb\ of integrated luminosity collected by the ATLAS
detector at the LHC. The differential cross-section has been measured
as a function of the prompt photon transverse energy between 45 and
400~\GeV, in the pseudorapidity ranges \etaone, \etatwo, \etathree\ and
\etafour. In general, good agreement
between the data and the NLO pQCD predictions is observed.
This measurement improves the precision and significantly extends the kinematic regime explored in the previous
measurement~\cite{InclusivePaper2010} and is consistent in
the region where the two measurements overlap. 

Over most of this extended kinematic range 
the experimental errors are smaller than the
theoretical ones. The large theoretical scale error limits the
discrimination between PDFs. Future measurements of
this process in finer pseudorapidity binning and those of the photon + jet system
should provide more insight into the PDF differences.

We thank CERN for the very successful operation of the LHC, as well as the
support staff from our institutions without whom ATLAS could not be
operated efficiently.

We acknowledge the support of ANPCyT, Argentina; YerPhI, Armenia; ARC,
Australia; BMWF, Austria; ANAS, Azerbaijan; SSTC, Belarus; CNPq and FAPESP,
Brazil; NSERC, NRC and CFI, Canada; CERN; CONICYT, Chile; CAS, MOST and
NSFC, China; COLCIENCIAS, Colombia; MSMT CR, MPO CR and VSC CR, Czech
Republic; DNRF, DNSRC and Lundbeck Foundation, Denmark; ARTEMIS, European
Union; IN2P3-CNRS, CEA-DSM/IRFU, France; GNAS, Georgia; BMBF, DFG, HGF, MPG
and AvH Foundation, Germany; GSRT, Greece; ISF, MINERVA, GIF, DIP and
Benoziyo Center, Israel; INFN, Italy; MEXT and JSPS, Japan; CNRST, Morocco;
FOM and NWO, Netherlands; RCN, Norway; MNiSW, Poland; GRICES and FCT,
Portugal; MERYS (MECTS), Romania; MES of Russia and ROSATOM, Russian
Federation; JINR; MSTD, Serbia; MSSR, Slovakia; ARRS and MVZT, Slovenia;
DST/NRF, South Africa; MICINN, Spain; SRC and Wallenberg Foundation,
Sweden; SER, SNSF and Cantons of Bern and Geneva, Switzerland; NSC, Taiwan;
TAEK, Turkey; STFC, the Royal Society and Leverhulme Trust, United Kingdom;
DOE and NSF, United States of America.

The crucial computing support from all WLCG partners is acknowledged
gratefully, in particular from CERN and the ATLAS Tier-1 facilities at
TRIUMF (Canada), NDGF (Denmark, Norway, Sweden), CC-IN2P3 (France),
KIT/GridKA (Germany), INFN-CNAF (Italy), NL-T1 (Netherlands), PIC (Spain),
ASGC (Taiwan), RAL (UK) and BNL (USA) and in the Tier-2 facilities
worldwide.

\bibliographystyle{model1-num-names}
\bibliography{InclusiveCrossSection2011}

\appendix

\section{Cross-section measurements}
\label{app:CStables}

Tables
\ref{tab:xsection_systematics_eta0}-\ref{tab:xsection_systematics_eta4}
list the values of the measured isolated prompt photon production
cross-sections, for the \etaone, \etatwo, \etathree\ and \etafour\
regions, respectively. The various systematic uncertainties
originating from the purity measurement, the photon selection and
identification efficiency and the luminosity are shown. In addition,
the correlated uncertainties between the efficiency and the purity
determination are propagated as such and included separately
($\sigma_{corr}$). The total uncertainty is the combination of the statistical and
systematic uncertainties (summed in quadrature), except for the
uncertainty on the luminosity.

\begin{table*}
  \caption{Measured isolated prompt photon cross-section for \etaonesh\ with statistical and systematic uncertainties. The total uncertainty includes both the statistical and all systematic uncertainties (summed in quadrature), except for the
uncertainty on the luminosity.}
  \label{tab:xsection_systematics_eta0}
  \centering
  \footnotesize
  \begin{tabular}{rrr|rrrrrr|r}
\hline\hline
$E_T^\mathrm{min}$ & $E_T^\mathrm{max}$ & $d\sigma / dE_T$ & $\delta_\mathrm{stat}$ & $\delta_\mathrm{yield}$ & $\delta_{\mathrm{efficiency}}$& $\delta_{\mathrm{corr}}$ & $\delta_\mathrm{unfolding}$ & $\delta_\mathrm{tot}$ & $\delta_\mathrm{lumi}$ \\
 $[\mathrm{GeV}]$ & $[\mathrm{GeV}]$ & $[\mathrm{pb/GeV}]$ & $[\mathrm{pb/GeV}]$ & $[\mathrm{pb/GeV}]$ & $[\mathrm{pb/GeV}]$ & $[\mathrm{pb/GeV}]$ & $[\mathrm{pb/GeV}]$ & $[\mathrm{pb/GeV}]$ & $[\mathrm{pb/GeV}]$\\
\hline
45 & 55 & 83.3 & 0.5 & 4.8 & 3.3 & 3.4 & 2.5 & 7.2 & 2.8 \\ 
55 & 70 & 32.7 & 0.3 & 1.8 & 1.2 & 1.2 & 1.0 & 2.7 & 1.1 \\ 
70 & 85 & 12.3 & 0.2 & 0.6 & 0.4 & 0.4 & 0.4 & 0.9 & 0.4 \\ 
85 & 100 & 5.3 & 0.1 & 0.2 & 0.2 & 0.2 & 0.2 & 0.4 & 0.2 \\ 
100 & 125 & 2.2   & 0.05 & 0.09 & 0.08 & 0.07 & 0.07 & 0.2 & 0.07 \\ 
125 & 150 & 0.80   & 0.03 & 0.03 & 0.03 & 0.02 & 0.03 & 0.06 & 0.03 \\ 
150 & 200 & 0.26   & 0.01   & 0.01 & 9\pcs{-3} & 7\pcs{-3} & 8\pcs{-3} & 0.02 & 9\pcs{-3} \\ 
200 & 400 & 2.8\pcs{-2} & 2\pcs{-3} & 2\pcs{-3} & 1\pcs{-3} & 4\pcs{-4} & 8\pcs{-4} & 3\pcs{-3} & 9\pcs{-4} \\ 
\hline\hline
\end{tabular}
\end{table*}

\begin{table*}
  \caption{Measured isolated prompt photon cross-section for \etatwo, uncertainties as in Table~\ref{tab:xsection_systematics_eta0}.}
  \label{tab:xsection_systematics_eta1}
  \centering
  \footnotesize
\begin{tabular}{rrr|rrrrrr|r}
\hline\hline
$E_T^\mathrm{min}$ & $E_T^\mathrm{max}$ & $d\sigma/ dE_T$ & $\delta_\mathrm{stat}$ & $\delta_\mathrm{yield}$ & $\delta_{\mathrm{efficiency}}$& $\delta_{\mathrm{corr}}$ & $\delta_\mathrm{unfolding}$ & $\delta_\mathrm{tot}$ & $\delta_\mathrm{lumi}$ \\
 $[\mathrm{GeV}]$ & $[\mathrm{GeV}]$ & $[\mathrm{pb/GeV}]$ & $[\mathrm{pb/GeV}]$ & $[\mathrm{pb/GeV}]$ & $[\mathrm{pb/GeV}]$ & $[\mathrm{pb/GeV}]$ & $[\mathrm{pb/GeV}]$ & $[\mathrm{pb/GeV}]$ & $[\mathrm{pb/GeV}]$\\
\hline
45 & 55 & 99.0 & 0.7 & 8.1 & 4.4 & 3.8 & 3.0 & 10.4 & 3.4 \\ 
55 & 70 & 38.9 & 0.3 & 3.0 & 1.7 & 1.2 & 1.2 & 3.9 & 1.3 \\ 
70 & 85 & 14.9 & 0.2 & 1.1 & 0.7 & 0.4 & 0.5 & 1.4 & 0.5 \\ 
85 & 100 & 6.3 & 0.1 & 0.4 & 0.3 & 0.1 & 0.2 & 0.6 & 0.2 \\ 
100 & 125 & 2.7 & 0.06 & 0.2 & 0.1 & 0.06 & 0.08 & 0.2 & 0.09 \\ 
125 & 150 & 1.0 & 0.03 & 0.06 & 0.04 & 0.02 & 0.03 & 0.1 & 0.03 \\ 
150 & 200 & 0.29 & 0.01 & 0.02 & 0.01 & 7\pcs{-3} & 9\pcs{-3} & 0.03 & 0.01 \\ 
200 & 400 & 3.2\pcs{-2} & 2\pcs{-3} & 3\pcs{-3} & 2\pcs{-3} & 9\pcs{-4} & 1\pcs{-3} & 4\pcs{-3} & 1\pcs{-3} \\ 
\hline\hline
\end{tabular}

\end{table*}

\begin{table*}
  \caption{Measured isolated prompt photon cross-section for \etathree, uncertainties as in Table~\ref{tab:xsection_systematics_eta0}.}
  \label{tab:xsection_systematics_eta3}
  \centering
  \footnotesize
  \begin{tabular}{rrr|rrrrrr|r}
\hline\hline
$E_T^\mathrm{min}$ & $E_T^\mathrm{max}$ & $d\sigma / dE_T$ & $\delta_\mathrm{stat}$ & $\delta_\mathrm{yield}$ & $\delta_{\mathrm{efficiency}}$& $\delta_{\mathrm{corr}}$ & $\delta_\mathrm{unfolding}$ & $\delta_\mathrm{tot}$ & $\delta_\mathrm{lumi}$ \\
 $[\mathrm{GeV}]$ & $[\mathrm{GeV}]$ & $[\mathrm{pb/GeV}]$ & $[\mathrm{pb/GeV}]$ & $[\mathrm{pb/GeV}]$ & $[\mathrm{pb/GeV}]$ & $[\mathrm{pb/GeV}]$ & $[\mathrm{pb/GeV}]$ & $[\mathrm{pb/GeV}]$ & $[\mathrm{pb/GeV}]$\\
\hline
45 & 55 & 41.9 & 0.4 & 4.6 & 3.1 & 1.2 & 1.3 & 5.8 & 1.4 \\ 
55 & 70 & 15.7 & 0.2 & 1.6 & 1.0 & 0.4 & 0.5 & 2 & 0.5 \\ 
70 & 85 & 6.4 & 0.2 & 0.5 & 0.4 & 0.2 & 0.2 & 0.7 & 0.2 \\ 
85 & 100 & 2.4 & 0.08 & 0.2 & 0.2 & 0.05 & 0.08 & 0.3 & 0.08 \\ 
100 & 125 & 1.0 & 0.04 & 0.07 & 0.08 & 0.02 & 0.03 & 0.1 & 0.03 \\ 
125 & 150 & 0.36 & 0.02 & 0.03 & 0.03 & 8\pcs{-3} & 0.01 & 0.05 & 0.01 \\ 
150 & 200 & 0.11 & 9\pcs{-3} & 0.01 & 7\pcs{-3} & 3\pcs{-3} & 4\pcs{-3} & 0.02 & 4\pcs{-3} \\ 
200 & 400 & 1.1\pcs{-2} & 1\pcs{-3} & 1\pcs{-3} & 8\pcs{-4} & 2\pcs{-4} & 3\pcs{-4} & 2\pcs{-3}& 4\pcs{-4} \\ 
\hline\hline
\end{tabular}

\end{table*}

\begin{table*}
  \caption{Measured isolated prompt photon cross-section for \etafour, uncertainties as in Table~\ref{tab:xsection_systematics_eta0}.}
  \label{tab:xsection_systematics_eta4}
  \centering
  \footnotesize
\begin{tabular}{rrr|rrrrrr|r}
\hline\hline
$E_T^\mathrm{min}$ & $E_T^\mathrm{max}$ & $d\sigma / dE_T$ & $\delta_\mathrm{stat}$ & $\delta_\mathrm{yield}$ & $\delta_{\mathrm{efficiency}}$& $\delta_{\mathrm{corr}}$ & $\delta_\mathrm{unfolding}$  & $\delta_\mathrm{tot}$ & $\delta_\mathrm{lumi}$ \\
 $[\mathrm{GeV}]$ & $[\mathrm{GeV}]$ & $[\mathrm{pb/GeV}]$ & $[\mathrm{pb/GeV}]$ & $[\mathrm{pb/GeV}]$ & $[\mathrm{pb/GeV}]$ & $[\mathrm{pb/GeV}]$ & $[\mathrm{pb/GeV}]$ & $[\mathrm{pb/GeV}]$ & $[\mathrm{pb/GeV}]$\\
\hline
45 & 55 & 68.9 & 0.6 & 7.6 & 3.8 & 3.9 & 2.1 & 9.6 & 2.3 \\ 
55 & 70 & 26.4 & 0.3 & 2.7 & 1.3 & 1.3 & 0.8 & 3.3 & 0.9 \\ 
70 & 85 & 10.0 & 0.2 & 0.9 & 0.5 & 0.5 & 0.3 & 1.2 & 0.3 \\ 
85 & 100 & 4.2 & 0.1 & 0.3 & 0.3 & 0.2 & 0.1 & 0.5 & 0.1 \\ 
100 & 125 & 1.7 & 0.06 & 0.1 & 0.1 & 0.08 & 0.05 & 0.2 & 0.06 \\ 
125 & 150 & 0.55 & 0.03 & 0.03 & 0.03 & 0.02 & 0.02 & 0.06 & 0.02 \\ 
150 & 200 & 0.17 & 0.01 & 0.01 & 0.01 & 6\pcs{-3} & 6\pcs{-3} & 0.02 & 6\pcs{-3} \\ 
200 & 400 & 1.2\pcs{-2} & 1\pcs{-3} & 6\pcs{-4} & 3\pcs{-3} & 3\pcs{-4} & 4\pcs{-4} & 3\pcs{-3} & 4\pcs{-4} \\ 
\hline\hline
\end{tabular}
\end{table*}

\onecolumn
\clearpage
\input{atlas_authlist}

\end{document}

%% file: definitions.tex
\newcommand{\etaonesh}   {\ensuremath{|\eta|<0.6}}

\newcommand{\etaone}   {\ensuremath{0.0\leq|\eta|<0.6}}
\newcommand{\etatwo}   {\ensuremath{0.6\leq|\eta|<1.37}}
\newcommand{\etathree}   {\ensuremath{1.52\leq|\eta|<1.81}}
\newcommand{\etafour}   {\ensuremath{1.81\leq|\eta|<2.37}}

\newcommand{\Etiso}   {\ensuremath{E_{\mathrm{T}}^{\mathrm{iso}}}}

\newcommand{\pythia} {\textsc{Pythia}}
\newcommand{\herwig} {\textsc{Herwig}}
\newcommand{\jetphox} {\textsc{Jetphox}}

\newcommand{\pcs}[1]{\ensuremath{\times10^{#1}}}

\newcommand{\efftrig}{\ensuremath{\varepsilon^{\rm trig}}}
\newcommand{\effreco}{\ensuremath{\varepsilon^{\rm reco}}}
\newcommand{\effid}{\ensuremath{\varepsilon^{\rm ID}}}

%% file: atlas_authlist.tex
\begin{flushleft}
{\Large The ATLAS Collaboration}

\bigskip

G.~Aad$^{\rm 48}$,
B.~Abbott$^{\rm 111}$,
J.~Abdallah$^{\rm 11}$,
A.A.~Abdelalim$^{\rm 49}$,
A.~Abdesselam$^{\rm 118}$,
O.~Abdinov$^{\rm 10}$,
B.~Abi$^{\rm 112}$,
M.~Abolins$^{\rm 88}$,
H.~Abramowicz$^{\rm 153}$,
H.~Abreu$^{\rm 115}$
% ...
\bigskip

% ...
$^{10}$ Institute of Physics, Azerbaijan Academy of Sciences, Baku, Azerbaijan\\
$^{11}$ Institut de F\'isica d'Altes Energies and Universitat Aut\`onoma  de Barcelona and ICREA, Barcelona, Spain\\
% ...
$^{48}$ Fakult\"{a}t f\"{u}r Mathematik und Physik, Albert-Ludwigs-Universit\"{a}t, Freiburg i.Br., Germany\\
$^{49}$ Section de Physique, Universit\'e de Gen\`eve, Geneva, Switzerland\\
% ...
$^{88}$ Department of Physics and Astronomy, Michigan State University, East Lansing MI, United States of America\\
% ...
$^{111}$ Homer L. Dodge Department of Physics and Astronomy, University of Oklahoma, Norman OK, United States of America\\
$^{112}$ Department of Physics, Oklahoma State University, Stillwater OK, United States of America\\
% ...
$^{115}$ LAL, Univ. Paris-Sud and CNRS/IN2P3, Orsay, France\\
% ...
$^{118}$ Department of Physics, Oxford University, Oxford, United Kingdom\\
% ...
$^{153}$ Raymond and Beverly Sackler School of Physics and Astronomy, Tel Aviv University, Tel Aviv, Israel\\
% ...
\end{flushleft}


%% file: InclusiveCrossSection2011.bbl
\begin{thebibliography}{24}
\bibitem[{Atl(eta2)}]{AtlasCoords}
\bibinfo{title}{ATLAS uses a right-handed coordinate system with its origin at
  the nominal interaction point (IP) in the centre of the detector and the
  $z$-axis along the beam pipe. The $x$-axis points from the IP to the centre
  of the LHC ring, and the $y$ axis points upward. Cylindrical coordinates
  $(r,\phi)$ are used in the transverse plane}, \bibinfo{year}{$\phi$ being the
  azimuthal angle around the beam pipe. The pseudorapidity is defined in terms
  of the polar angle $\theta$ as $\eta=-\ln\tan(\theta/2)$.}
\end{thebibliography}
